\documentclass[preprint,showpacs,preprintnumbers,amsmath,amssymb,superscriptaddress,nofootinbib,prb]{revtex4}
\usepackage{bm}
\usepackage{graphicx}

\begin{document}
\title[Spin-concentration dependence of the freezing temperature in metallic spin-glasses]{Metallic spin-glasses beyond mean-field: an approach to the impurity-concentration dependence of the freezing temperature}

\author{E Cuervo-Reyes}

\affiliation{Laboratory of Inorganic Chemistry, Solid State, Swiss Institute of Technology (ETH),   CH-8093 Z\"urich}
\email{cuervo@inorg.chem.ethz.ch}

 \begin{abstract}
A relation between the freezing temperature ($T^{}_{\rm g}$) and the exchange couplings  ($J^{}_{ij}$) in metallic spin-glasses is derived, taking the spin-correlations ($G^{}_{ij}$) into account. This approach does not involve a disorder-average. The expansion of the correlations to first order in $J^{}_{ij}/T^{}_{\rm g}$ leads to the molecular-field result from Thouless-Anderson-Palmer. Employing the current theory of the spin-interaction in disordered metals, an equation for $T^{}_{\rm g}$ as a function of the concentration of impurities is obtained, which reproduces the available data from {\sl Au}Fe, {\sl Ag}Mn, and {\sl Cu}Mn alloys well. 
\end{abstract} 
\maketitle

\section{Introduction}
Spin glasses are disordered, and partially frustrated magnetic systems with a critical (freezing) temperature $T^{}_{\rm g}$, below which the spins are  frozen in random orientations without a conventional order\cite{General,General2,Book} \footnote{In contrast to what the word ``glass'' suggests, spin-freezing is a true phase transition.}.  Diluted alloys of transition-metal  impurities  in a noble-metal matrix (e.g., {\sl Au}Fe, {\sl Ag}Mn, and {\sl Cu}Mn) are typical examples, and are known as canonical spin-glasses (CSGs)\cite{Book}. These systems have been the subject of intense research during the last five decades, and many of their fundamental characteristics are now well understood. However, some questions still lack a satisfactory answer.      

A fundamental, and  not yet fully solved problem is the dependence of $T^{}_{\rm g}$ on the composition of the alloy; i.e., how $T^{}_{\rm g}$ varies, as we change the concentrations ($c$, and $c^{}_i$) of magnetic, and non magnetic impurities. Valuable experimental data have been collected by several researchers\cite{Book,Myd,Myd2,Cowen,VS}; yet, the observed behaviour is only partially understood. A primary question is: can we derive a formula that describes the experiments coherently, in a  quantitative manner? This is, actually, a two-fold problem.  In CSGs, the spins are coupled by means of the conduction electrons. The exact form of this interaction, as a function of the distance between the spins ($r^{}_{ij}$), is only known for the ideal case of two spins in an otherwise clean metal. There, it is given by the Ruderman-Kittel-Kasuya-Yosida\cite{RKKYI} (RKKY) formula  $J^{}_{ij}\propto r^{-3}_{ij} \cos(2k^{}_{\rm F} r^{}_{ij})$, $k^{}_{\rm F}$ being the Fermi wave-vector.  Randomly distributed, magnetic and non-magnetic impurities scatter the conduction electrons, and therefore, modify this $J(r^{}_{ij})$ dependence. Thus, in order to describe $T^{}_{\rm g}(c,c^{}_i)$  one needs to solve two general problems: (1) the form of the effective couplings $J^{}_{ij}$ in metals with impurities, and (2) the functional form $T^{}_{\rm g}[J^{}_{ij}]$.

 The effective interaction, in the presence of disorder, has been studied by many authors\cite{D1,D2,D3,D4,D5,Geldart,Larsensat,LarsenKane,KaneTheo1,KaneTheo2}. Analytical results have been obtained in some asymptotic limits, which can be employed with the assistance of interpolating forms. Further advances, within analytical methods, seem quite challenging. The present work  focuses on solving the second problem; finding the functional dependence $T^{}_{\rm g}[J^{}_{ij}]$. 

 Perhaps the best known approximation to $T^{}_{\rm g}[J^{}_{ij}]$ is $T^{}_{\rm g}\propto\sqrt{\frac{1}{N}\sum^{}_{ij} J^{2}_{ij}}$, from the Edwards-Anderson model\cite{EA}, and also derived by Sherrington\cite{Sher}, neglecting the correlations between the orientation of the spins and the couplings. Here on, I shall refer to this result as  MFF (mean-field-formula); similarly, the general term mean-field will be abbreviated as MF. This same dependence (to within a multiplicative constant) was obtained by Thouless, Anderson, and Palmer (TAP), following a rather different approach\cite{TAP}. They extracted  $T^{}_{\rm g}$ from the eigenvalue distribution of the interaction matrix  $J^{}_{ij}$, corrected by a reaction field, under the assumption of Gaussian fluctuations in the local fields.  The  MFF  possesses some desired features. It gives a finite  $T^{}_{\rm g}$ in the limit of infinite range interactions (clean limit), and, if the shortest  distance between spins is taken as $\propto c^{-1/3}_{}$ (proportional to the typical distance), one obtains  $T^{}_{\rm g}\propto c$, in agreement with scaling arguments. However,  there has been no conclusive evidence to show that the MFF can also account for dirt-related effects, using physical parameters with realistic values. Moreover, neither the absence of local correlations between the spins, nor the statistical independence of spins and couplings seems to be justified\cite{Geldart2}. The assumption of Gaussian fluctuations  in metallic spin-glasses is also debatable\cite{Vug}.
 
 A different phenomenological relation $T^{}_{\rm g}\sim \frac{1}{N}\sum^{}_{ij} |J^{}_{ij}|$ was proposed by Shegelski and Geldart (SG)\cite{Geldart2} (see next section), and with it they obtained an overall acceptable fit to the experiments, at moderate $c$. They could also describe, in a semi-quantitative manner, the changes in $T^{}_{\rm g}$ caused by the addition of non-magnetic impurities. Unfortunately, this formula predicts an asymptotic dependence  $T^{}_{\rm g}\propto -c\ln c$, when $c\rightarrow 0$, and an infinite $T^{}_{\rm g}$ in the clean limit (in the absence of electron mean-free-path effects). 
  
Despite the continuous research in this field\cite{epl0,epl}, the theoretical and quantitative description of experiments remain as challenging as two decades before\footnote{The random-local-field-approach\cite{Vug}, by Vugmeister {\it et.al.}, does not allow an analytical solution with the RKKY interaction.}.  Simplified spin models exist that allow more exact solutions; but these cannot be generalized to realistic interactions. Since the exact form of the electron-mediated (RKKY) interaction is only known in some asymptotic limits, it is  difficult to judge whether the discrepancies between experiments and theory are only due to the MF approximations, or also due to the chosen interpolations for  $J^{}_{ij}$. Another reason to look for solutions beyond the MFF  is that fluctuations are very  important at the critical point, and therefore, we may expect (perhaps large) deviations from the MF predictions. In fact, the relevant number of spatial dimensions in CSGs is $d=3$, which is rather close to the lower critical dimension ($2<d^{}_l\leq 3$). The upper critical dimension (where the standard MF-approximation becomes exact) is $d^{}_u=6$\cite{BMY}.

In this paper, an alternative equation for the freezing temperature in metallic spin-glasses is derived. First, it is obtained as a functional of the interaction matrix $T^{}_{\rm g}[J^{}_{ij}]$ (in  section \ref{theo}), where we also discuss its relation to previous approaches.  $T^{}_{\rm g}[J^{}_{ij}]$ is rewritten in terms of the distance dependence of the interaction and the spin-correlations, in sections \ref{shells} and \ref{G}. These are the main results of the present work. The rest of the paper (sections \ref{Calc}, and \ref{Exp}) is dedicated to the comparison with experiments. Conclusions are presented in section \ref{End}.

\section{The functional $\boldsymbol{ T^{}_{\rm g}[J^{}_{ij}]}$}\label{theo}

The system of interacting spins is represented by  the classical Hamiltonian
\begin{equation}
H=-\sum_{i=1}^N\sum_{j>i}^N J^{}_{ij}S^{}_iS^{}_j\; .\label{Hamil}
\end{equation}
\noindent The spins $S^{}_i$s can be Ising variables, or vectors of unit length; the actual length can be included by rescaling the coupling constants. The $J^{}_{ij}$s are  symmetrically distributed around zero, which follows from the sampling of the RKKY interaction, according to randomly located spins, at low concentrations.

 For $T$ just below $T^{}_{\rm g}$, the self-consistency MF-equations (for a given spatial arrangement of the impurities) are
\begin{equation}
\langle S^{}_i\rangle=\beta \sum^{}_j J^{}_{ij} \langle S^{}_j\rangle \;,\label{MF0}
\end{equation}

\noindent  where $\beta=T^{-1}_{}$, (with the Boltzmann constant  set to $k^{}_{\text B}=1$).When  (\ref{MF0}) is multiplied by $\langle S^{}_i\rangle$, and averaged over sites, it becomes
  
\begin{equation}
\frac{1}{N}\sum_i |\langle S^{}_i\rangle|^2_{}= \frac{\beta}{N}\sum_i\sum^{}_j J^{}_{ij} \langle S^{}_i\rangle\langle S^{}_j\rangle \;.\label{MF1}
\end{equation}  

\noindent The lhs of  (\ref{MF1}) can be  identified as the Edwards-Anderson (EA) order parameter\cite{EA}
\begin{equation}
q^2_{}=\frac{1}{N}\sum_i|\langle S^{}_i\rangle|^2_{}\;,\label{EAOP}
\end{equation}  
\noindent which is a global quantity, and acquires non vanishing values for $T<T^{}_{\rm g}$. In order to obtain a useful self-consistency equation for $q^2_{}$, and therefore, a solution for $T^{}_{\rm g}$, one needs to express the rhs of   (\ref{MF1}) in terms of $q^2_{}$.  Following the symmetry in the distribution of couplings,  the average over pairs of sites with $i\neq j$ satisfy

\begin{subequations}
\begin{equation}
\overline{\langle S^{}_i\rangle\langle S^{}_j\rangle}=0\; , 
\end{equation}
\begin{equation}
\overline{\langle S^{}_i S^{}_j\rangle}=0\; ,  
\end{equation}
\noindent and 
\begin{equation}
\overline{J^{}_{ij}\langle S^{}_i S^{}_j\rangle} >0\; ,\label{cor1}
\end{equation}
\noindent for all $T$; whereas 
\begin{equation}
\overline{J^{}_{ij}\langle S^{}_i\rangle\langle S^{}_j\rangle} >0\; ,\label{cor2}
\end{equation}
\end{subequations}
\noindent for $T<T^{}_{\rm g}$, and it is zero otherwise. That (\ref{cor1}) and (\ref{cor2}) are positive follows from the fact that the apparent randomness of the spin orientations is controlled by the distribution of couplings. The system minimizes the energy by orienting the spins (in average) according to the strongest couplings, as long as the frustration allows it. For symmetric distributions of couplings (with zero as a dominant mode), only about half of the system is frustrated\cite{vH0,vH}. The averages (\ref{cor1}) and (\ref{cor2}) would be zero only for a fully frustrated system (not having a finite temperature transition).  Taking this into account, and noticing that  $|\langle S^{}_i\rangle\langle S^{}_j\rangle|$ should grow, on average,  as $q^2_{}$, SG proposed\cite{Geldart2} that  $\frac{1}{N}\sum_i\sum^{}_j J^{}_{ij} \langle S^{}_i\rangle\langle S^{}_j\rangle$ can be estimated with  $\frac{1}{N}\sum_i\sum^{}_j |J^{}_{ij}|q^2_{}$, to within a constant factor of order one, which reflects the partial frustration.  Vugmeister {\it et al}\cite{Vug3} pointed out that, for the RKKY interaction (where $|J^{}_{ij}|\propto r^{-3}_{ij}$), this estimate diverges  in the absence of a long distance cut-off. The origin of the divergence, in this otherwise-reasonable approach, has not been discussed up to date.  As we will see immediately, understanding this detail is the key for finding a formula for $T^{}_{\rm g}$ beyond the MFF.
 
 Thermal and spacial fluctuations are inherent to thermodynamic systems at $T>0$, and they play an important role in the vicinity of critical points. For example, in a simple ferromagnetic system (where all couplings have positive sign) the probability that two spins point in the same direction decreases as we look at pairs of spins which are further apart. In other words, spin-correlations decay with the distance.  If we wish to estimate the double sum in  (\ref{MF1}), we must take this into account. The direction of the spin is rather determined by its closest environment (strongest couplings), and spins further apart have (on average) less influence. The probability of finding a pair of spins (at sites $i$, and $j$) pointing according to the direct coupling $J^{}_{ij}$ decreases as the distance between the spins becomes larger (because of the larger cloud of fluctuating spins that lies in between).   Therefore, it is reasonable to propose a slightly different estimate for the double sum, by introducing the correlation matrix $G^{}_{ij}$; i.e.,
\begin{equation}
\frac{1}{N}\sum_i\sum^{}_j J^{}_{ij} \langle S^{}_i\rangle\langle S^{}_j\rangle\rightarrow\frac{1}{N}\sum_i\sum^{}_j J^{}_{ij}G^{}_{ij}q^2_{} \;.\label{DS}
\end{equation}       
\noindent The values of the matrix terms $G^{}_{ij}$ cannot be calculated exactly. However, their average coarse grained behaviour as a function of the distance (see later in section \ref{G}), will be enough for arriving at an analytically amenable formula  for $T^{}_{\rm g}$. The factor that multiplies $q^2_{}$ is strictly positive, and  independent of the specific realization of the random site distribution. From (\ref{MF1}) and (\ref{DS}) one obtains 
\begin{equation}
T^{}_{\rm g}\propto \frac{1}{N}\sum_{ij}^{} J^{}_{ij}G^{}_{ij}\; ,\label{TGex}
\end{equation}
\noindent  to within a numerical constant.  The fall-off of the correlations ensures that $T^{}_{\rm g}<\infty$. Note that since $\frac{1}{N}\sum^{}_{ij} |J^{}_{ij}|$ diverges logarithmically, any power-law decay suffices to cancel this singularity.

At this point, we may regard (\ref{TGex}) as a phenomenological solution to $T^{}_{\rm g}[J^{}_{ij}]$, as a correction to the SG approach. However, (\ref{TGex}) can also be derived following a more fundamental route, by analysing the self-response of the spins on the high-$T$ side of the transition. Presume that some spins acquire a non-vanishing,  small thermal average.  This perturbation of the paramagnetic state would change the local fields by $ \delta h^{}_j= \sum_k J^{}_{jk}\delta S^{}_k$. $\delta h^{}_j$ would affect the neighbouring spins, and these, their neighbours, and so on. According to the linear-response theory, this far-reaching effect is given, at any point, in terms of the spin-spin correlations $G^{}_{ij}\equiv\langle S^{}_iS^{}_j\rangle$, as  
\begin{equation}
\delta S^{}_i=\beta \sum^{}_j G^{}_{ij}\delta h^{}_j\; .\label{LR0}
\end{equation}     
\noindent Now, one should check whether non-vanishing $\delta S^{}_i$s can be maintained  by the self-induced  fields; i.e., whether the system of equations
\begin{equation}
\delta S^{}_i=\beta \sum^{}_{jk} G^{}_{ij}J^{}_{jk} \delta S^{}_k \; \label{LR}
\end{equation} 
\noindent has non-trivial solutions. A non-zero solution, for a number of spins of order $N$, would signal an instability in the paramagnetic phase in the thermodynamic limit.  One difficulty with this task is that the matrix $Q^{}_{ik}\equiv\beta\sum^{}_{j} G^{}_{ij}J^{}_{jk}$ (whose elements decay with the distance) has localized, and delocalized eigenvectors, whereas the occurrence of the phase transition is related to delocalized states\cite{Vug,Vug3,Vug4,Evan} (the transition takes place when the mobility edge between delocalized and localized eigenvalues reaches the value of one). $Q^{}_{ik}$ has the following properties. (1) The off-diagonal terms have zero average and finite variance $\overline{Q^{2}_{ik}}=\beta^2_{}\sum_j \overline{G^{2}_{ij}} \,\overline{J^{2}_{jk}}<\infty$. This derives from the fact that (for $i\neq k$) each of the correlation terms $G^{}_{ij}$ and the multiplying  coupling $J^{}_{jk}$, correspond to two different spin-pairs; therefore they are statistically independent. (2) The diagonal elements are positive, because they are proportional to  the effective energy per site $T Q^{}_{ii}=-\epsilon^{}_i(T)\equiv\sum_j G^{}_{ij}J^{}_{ji}$ (similar to (\ref{cor1})). In thermal equilibrium,   $\epsilon^{}_i(T)\approx \epsilon(T)=- \frac{1}{N}\sum_{ij} G^{}_{ij}J^{}_{ji}<0$. Then, $Q^{}_{ik}$ can be written as an average scalar matrix ($\beta |\epsilon|\delta^{}_{ik}$, where $\delta^{}_{ik}$ is the identity matrix), plus a fluctuation matrix ($\beta O^{}_{ik}$),
\begin{equation}
Q^{}_{ik}=\beta |\epsilon|\delta^{}_{ik}+\beta O^{}_{ik}  \;.
\end{equation}

   To get the global information out of this system of equations, it is useful to introduce an EA type of parameter. Multiplying (\ref{LR}) by $\delta S^{}_i$, and averaging over sites,

\begin{eqnarray}
q^2_{}&=&\frac{1}{N}\sum_i |\delta S^{}_i|^2_{}\; \nonumber\\
      &=&\beta|\epsilon| q^2_{} +\beta \frac{1}{N}\sum_{ik} O^{}_{ik} \delta S^{}_i\delta S^{}_k\; .\label{EALR}
\end{eqnarray}
\noindent The idea behind this is that localized eigenstates, where only a finite number of spins have non-zero values, will have vanishing $q^2_{}$ in the thermodynamic limit. So, one can screen out those contributions.  The double sum in  (\ref{EALR}) can be rewritten as 
\begin{equation}
\beta \frac{1}{N}\sum_{ik} O^{}_{ik} \delta S^{}_i\delta S^{}_k=\frac{\beta}{\sqrt{N}}\sum_k \lambda^{}_k \delta S^{2}_{\lambda^{}_k }\label{vanish}
\end{equation}
\noindent in terms of the eigenvalues ($\lambda^{}_k$) of the rescaled matrix $O'_{ik}\equiv N^{-1/2}_{}O^{}_{ik}$, which are symmetrically distributed around zero, and satisfy   $|\lambda^{}_k|<\infty$ for any $N$\cite{Wigner}. $\delta S^{2}_{\lambda^{}_k }$ is the square of the projection of the $N$-spin vector $\vec{\delta S}=$ ($\delta S^{}_1$, $\delta S^{}_2$, $\ldots$, $\delta S^{}_N$) on the $k$-th eigenvector of $O'_{ik}$. Although the exact form of a delocalized eigenvector is unknown, it can be formally substituted in  (\ref{EALR}). This gives   
\begin{equation}
q^2_{}=\beta|\epsilon| q^2_{} +\frac{\beta}{\sqrt{N}}\lambda^{}_{\text{deloc}}\; . \label{MFdloc}
\end{equation}
\noindent Now, the limit $N\rightarrow\infty$ is taken; the second term in the  rhs  vanishes, and  (\ref{MFdloc}) becomes 
\begin{equation}
q^2_{}=\beta|\epsilon| q^2_{} \; .
\end{equation}
\noindent This gives us again (\ref{TGex}), as a condition for  $q^2_{}$ acquiring non-zero values (this time with a known numerical prefactor). Thus, provided $q^2_{}>0$ is a signature of the ordered phase\footnote{The EA parameter is not a true order parameter because it does not single out the ergodic component that the system picked up in the lower symmetry phase\cite{vEvH}. It does not even differ between spin-glass and conventional order (e.g., ferromagnetic order). Nevertheless, a non-zero $q$ does indicate that a macroscopic number of spins has acquired a non-zero thermal average, which allows us to employ it as a signature of the transition. The chosen distribution of couplings excludes the possibility of any conventional order.}, $T^{}_{\rm g}$ is obtained from the equation
\begin{equation}
T^{}_{\rm g}=|\epsilon(T^{}_{\rm g})|\; .\label{main}
\end{equation}
\noindent This proves the correctness of the ansatz (\ref{DS}).  The energy per spin is 
\begin{eqnarray}
\frac{\langle H\rangle}{N} & = & -\frac{1}{2N}\sum_{i,j}^{} J^{}_{ij}\langle S^{}_i S^{}_j\rangle \nonumber\\
  & = & -\frac{1}{2N}\sum_{i,j}^{} J^{}_{ij}\left(G^{}_{ij}+\langle S^{}_i\rangle\langle S^{}_j\rangle\right)\nonumber\\
  &= &  -\frac{1}{2N}\sum_{i,j}^{} J^{}_{ij}G^{}_{ij}\left(1+q^2_{}\right)\; ,\nonumber\\
  & = & \frac{\epsilon(T)}{2}\left(1+q^2_{}\right) .\label{ener}
\end{eqnarray}
If $G^{}_{ij}$ is expanded to the lowest order perturbation in $\beta J^{}_{ij}$ (i.e., $[G^{-1}_{}]^{}_{ij}=\delta^{}_{ij}-\beta  J^{}_{ij}$\cite{BM}, and consequently,   $G^{}_{ij}=\delta^{}_{ij}+\beta  J^{}_{ij}$),  (\ref{main}) reduces to the TAP result  $T^{}_{\rm g}=\sqrt{(1/N)\sum_{ij} J^{2}_{ij}}$. This quantity, as well as $|\epsilon(T)|$, does not depend on the specific configuration\cite{LD,LD2}. Sometimes\cite{General2} it is assumed that $\sum_{j} J^{2}_{ij}$ is also independent of the site $i$, implying that   $\sum_{j} J^{2}_{ij}\equiv (1/N)\sum_{ij} J^{2}_{ij}$. However, this is not exactly true.

 The spin-freezing occurs, as with any other magnetic transition, when the energy gained by ordering overcomes the loss of entropy. Hence, it seems natural that $T^{}_{\rm g}$ scales with the interaction energy per spin. The correlations between the couplings and the relative orientation of the spins  play a central role in the derivation of  (\ref{main}).  In the simple local-MF-approach, the total energy per spin
 \begin{equation}
 E^{}_{\rm MF}=-\frac{1}{2N}\sum^{}_{ij} J^{}_{ij}\langle S^{}_{i}\rangle\langle S^{}_{j}\rangle\; \label{ave1}
 \end{equation} 
 \noindent would be zero, if one assumes that the $J^{}_{ij}$s and the orientations of the spins are independent, which  leaves no reason for the ordering.   To get around this problem, Sherrington\cite{Sher} iterated  (\ref{MF0}) before performing a disorder average ($[\cdot]^{}_{\rm av}$),  and obtained an energy equivalent to
 \begin{eqnarray}
  E^{}_{\rm MF}&=&-\frac{1}{2N}\beta\sum^{}_{ijl} [J^{}_{ij}J^{}_{jl}\langle S^{}_{i}\rangle\langle  S^{}_{l}\rangle]^{}_{\rm av}\; ,\nonumber\\
   &=&-\frac{1}{2}\beta\sum^{}_{j}[J^{2}_{ij}]^{}_{\rm av} q^2_{}<0\;. \label{ave2}
 \end{eqnarray}

 \noindent  Equations (\ref{ave1}) and (\ref{ave2}) give  two different values for the energy, but no transparent interpretation of the physics behind this. Actually, $E^{}_{\rm MF}\equiv 0$ is the trivial $T\rightarrow\infty$ limit (no correlations), and  (\ref{ave2}) turns out to be the freezing contribution in   (\ref{ener}) after expanding $G^{}_{ij}$ to first order in $\beta J^{}_{ij}$. Note that in this order of approximation $|\epsilon|\approx \beta^{}_{}\frac{1}{N}\sum^{}_{ij} J^{2}_{ij}$. Similarly,  the  MFF can be rephrased as (\ref{main}), with $|\epsilon|$ replaced by its high-$T$ form, extrapolated down to $T^{}_{\rm g}$; i.e. $T^{}_{\rm g}=\beta^{}_{\rm g}\frac{1}{N}\sum^{}_{ij} J^{2}_{ij}$. The resulting dependence of $T^{}_{\rm g}$ on the interaction range 
is quite different to the one given by (\ref{main}).
 
The bare MF-approximation changes the properties of the matrix $Q^{}_{il}$, by replacing it with $J^{}_{il}$. Actually, $G^{}_{ij}$, whose values (and sign) fluctuate form site to site,  is replaced with a trivially self-averaging quantity (the identity matrix). In such situations,  resorting to  averages over different realizations of the disorder  could be a choice in order to recover disorder-independent equations. However, there has not been a unique rule of how to perform this average\cite{General2,EA,Sher}, and it is not clear whether the  essential features of the actual system are always preserved.    In contrast to thermal averages, the configurational average does not represent a physical process. The  physical system (with quenched positional disorder) does not mutate; it remains in one configuration, yet showing sample-independent properties. Since the free energy ($F$) should not depend on the disorder, a suitable implementation of the formal average $[F]^{}_{\rm av}$   could provide us with more amenable equations that still  represent the  system  correctly\cite{General2}. However, much care must be taken when performing disorder averages of other physical quantities, or when changing the sequence in which different  averages/sums are performed. The average over disorder does not necessarily make the physics more transparent.

Based on the reproducibility and the sharpness of the transition in metallic spin-glasses, it seems natural to think that  there could be an analytical approach within which $T^{}_{\rm g}$ can be obtained from a single configuration\cite{vH0,vH}. To this end, we would have to describe the desired properties in terms of global quantities (either extensive variables, or their densities), so that the fluctuations of the local quantities are washed out by the mere definition of these global quantities. The present derivation of (\ref{main}) constitutes an example of such an approach. 

\section{From a sum to an integral}\label{shells}     
In order to put  (\ref{main}) in a more amenable form,  the sum over sites is divided into  concentric shells of thickness ${\rm d}  r$, with $k^{-1}_{\rm F}\ll {\rm d}  r\ll \Lambda$. $\Lambda$ is the range of the interaction (to be specified later). Note that, on account of the oscillating sign  of the couplings (originating from the factor $\cos(2k^{}_{\rm F}r^{}_{ij})$), every shell contains a good sample of the distribution. Then  
 \begin{equation}
T^{}_{\rm g}= \int^{\infty}_{r^{}_0} \overline{G(r)J(r)}{\rm d} N(r)\; , \label{I1}
 \end{equation}  
\noindent where $r^{}_0$ is the shortest distance cut-off (to be specified later), and ${\rm d} N(r)$ is the number of spins in the layer. The argument $r$, indicates that the average over sites is constrained to pairs with $r\leq r^{}_{ij}< r+ {\rm d}  r$.  As half of the bonds are frustrated on average\cite{vH},  $\overline{G(r)J(r)}=\frac{1}{2}\overline{|G(r)||J(r)|}$. This correlated average is now required, whereas the information available (see later) concerns the properties of $\overline{|G(r)|}$, and $\overline{|J(r)|}$. There are, at least, three arguments supporting the replacement of $\overline{|G(r)||J(r)|}$, with $\overline{|G(r)|}\;\overline{|J(r)|}$. 
 
 First, we should notice that smallest $|G_{ij}|$s do not always correspond to smallest $|J_{ji}|$s. Take as an example any pair of spins for which $\cos(2k^{}_{\rm F} r^{}_{ij})=0$. The coupling $J^{}_{ji}$ is zero, and $G^{}_{ij}$ is most probably different from zero due to indirect correlations. The opposite example is also possible: $J^{}_{ji}\ne 0$, and $G^{}_{ij}=0$  because of the frustration. Second, one can easily show that $\overline{|G^{}_{ij}||J^{}_{ji}|}$ becomes identical to  $\overline{|G^{}_{ij}|}\;\overline{|J^{}_{ji}|}$,  if one replaces  $\cos(2k^{}_{\rm F} r^{}_{ij})$ with $\frac{1}{2}(\xi^{}_i\eta^{}_j+\xi^{}_j\eta^{}_i)$, where the $\xi$s, and $\eta$s are independent random variables taking values $\pm 1$ with equal probability.  This modification  preserves the most important characteristics of the distribution of $\cos(2k^{}_{\rm F} r^{}_{ij})$\cite{General,General2,vH0,vH}: (i) the random variables  $\frac{1}{2}(\xi^{}_i\eta^{}_j+\xi^{}_j\eta^{}_i)$ are weakly correlated and oscillate between $-1$ and $+1$; (ii) their probability distribution is symmetrical with respect to zero; and (iii) the  values around zero are twice as probable as the values close to the extremes. A third fact to consider is that  the correlations between the decays of $|G(r)|$ and  $|J(r)|$, which result from the dimensionality of the space, are not affected as long as the radial integral is done on the product $\overline{|G(r)|}\;\overline{|J(r)|}$.  Following the above reasoning, we may expect that 
\begin{equation}
T^{}_{\rm g}=\frac{1}{2} \int^{\infty}_{r^{}_0} \overline{|G|}\; \overline{|J|}{\rm d}N(r) \label{CC0}
 \end{equation}  
 \noindent is a very good approximation to (\ref{I1}).     

 \section{The critical behaviour of $\boldsymbol{\overline{|G(r)|}}$}\label{G}
 
 Experiments and Montecarlo simulations\cite{Crit,Crit2} have agreed that the non-linear susceptibility $\chi_{\rm nl}^{}\equiv\frac{1}{N}\sum^{}_{ij}\langle S^{}_iS^{}_j\rangle^2_{}$ diverges at $T^{}_{\rm g}$, and that the  spin-glass correlation $G^{}_{\rm SG}(r)\equiv\overline{\langle S(r)S(0)\rangle^2_{}}$ has a critical falloff $G^{}_{\rm SG}(r)\propto  (r^{}_c/r)^{1+\eta}_{}$, for $r>r^{}_c$. $r^{}_c\sim c^{-1/3}_{}$ is the typical spin distance. As the pair correlations $G^{}_{ij}$ are bounded (i.e., $|\langle S^{}_iS^{}_j\rangle|\leq 1$), the double inequality 
 \begin{equation}
 G^{}_{\rm SG}(r)\leq \overline{|G|}\leq \sqrt{G^{}_{\rm SG}(r)}\label{Ineq}
 \end{equation}
 \noindent  holds for any $r$. This double inequality implies that $\overline{|G|}$  also has an infinite range at $T=T^{}_{\rm g}$, and that
 \begin{equation}
 \overline{|G|}\propto \left(\frac{r^{}_c}{r}\right)^{1+\eta'}_{}\; , \label{Gabs}
 \end{equation}
 \noindent with 
 \begin{equation}
 -1<\frac{\eta -1}{2}\leq\eta'\leq \eta\; .\label{eta}
 \end{equation} 
\noindent  Whether $\eta'$ is a constant (a true critical exponent) or $r$-dependent, is to be tested by means of Montecarlo simulations. From a geometrical perspective, it seems quite reasonable that $ \overline{|G|}\propto\sqrt{G^{}_{\rm SG}(r)}$, to within a numerical factor of order (and smaller than) one.  Later this will be useful in order to get an analytical expression for  $T^{}_{\rm g}$.

A finite cusp  in the linear susceptibility at $T=T^{}_{\rm g}$ is another characteristic of spin glasses; and it implies that the correlation function $\overline{G}$ has a finite range. The long tail for $\overline{|G|}$ does not contradict this observation. $\overline{G}$ does not show a critical decay, because of the fluctuating sign for $G^{}_{ij}$. The  double sums $\sum_{ij} |G^{}_{ij}|^2_{}$ and $\sum_{ij} |G^{}_{ij}|$ grow faster than the system size ($N$), whereas   $\sum_{ij} G^{}_{ij}\propto N$.

\section{An equation for $\boldsymbol{T^{}_{\rm g}(c,c^{}_i)}$}\label{Calc}
\subsection{The effective interaction}\label{jeff}
A complete solution for $T^{}_{\rm g}$, as a function of system parameters, requires an expression for the effective interaction, to be substituted in  (\ref{CC0}). We will shortly  review the state of the art on this subject; then, the interpolation model will be defined. Detailed  information can be found in the works of S-G\cite{Geldart,Geldart2}. 

The strength of the effective interaction (leaving out the oscillating factor) can be written as
\begin{equation}
J(r)=\frac{9\pi J^2_{sd} S^2_{}(2l+1)^2_{}}{T^{}_{\rm F}(2k^{}_{\rm F}r)^3_{}}h(r)
\end{equation} 
\noindent where $l=2$ (for $3d$-transition metal impurities), $S$ is the length of the spins, $T^{}_{\rm F}$ is the Fermi energy (or temperature, remember that we took $k^{}_{\rm B}=1$), and $J^{}_{sd}$ is the $s$-$d$ exchange coupling. $h(r)$ is a function that  depends strongly on the composition, having the asymptotic forms
\begin{equation}
h(r)\sim\left\{\begin{array}{lcl}
      1 & , & r^{}_0 < r < \Lambda^{}_{+} \\
      \left[I^{}_0(r)+I^{}_1(r)\right]^{1/2}_{} & , & \Lambda^{}_{+} < r < \Lambda^{}_T \\
     \text{e}^{-r/\Lambda^{}_T}  & , & \Lambda^{}_{T} < r
\end{array}  \right.
\end{equation}
  
\noindent where $I^{}_0\sim\frac{3}{16}$, and  $I^{}_1$ has a dominant decay $\sim \text{e}^{-2r/\Lambda^{}_{+}}_{}$. 

$\Lambda^{}_{T}=\left(\lambda\lambda^{}_{T}/3 \right)^{1/2}_{}$ is the range of the interaction at finite temperature, resulting from the cooperative effect of scattering and thermal diffusion. $\lambda=(\lambda^{-1}_{sd}+\lambda^{-1}_{i})^{-1}_{}$ is the total electron mean free path.  $\lambda^{}_{sd}$ is the mean free path due to $sd$ scattering by magnetic impurities, while $\lambda^{-1}_{i}$ accounts for the collisions with non-magnetic impurities. $\lambda^{}_{T}=T^{}_{\rm F}/(\pi k^{}_{\rm F} T)$ is the thermal coherence length.

$\Lambda^{}_{+}$ is a complex length scale, an effective range. When only magnetic impurities are present, $\Lambda^{}_{+}=\lambda$. When non-magnetic impurities are added, so that the total resistivity from both types of impurities satisfy $\rho=\rho^{}_{sd}+\rho^{}_{i}\gg \rho^{}_{sd}$, then $\Lambda^{}_{+}\sim  \sqrt{\lambda^{}_{sd}\lambda/2}>\lambda$. The physics behind this is that when electrons are scattered by non-magnetic impurities, they may not lose the spin-information, which can then be transferred to other spins. The effective range does not decrease as fast as $\lambda$. In general $\lambda\leq\Lambda^{}_{+}<\Lambda^{}_{T}$.

In the following, the  interpolation forms
\begin{subequations}
\begin{eqnarray}
h(r) & = & \left[\frac{13}{16}\text{e}^{-2r/\Lambda^{}_{+}}_{}+\frac{3}{16}\text{e}^{-2r/\Lambda^{}_{T}}_{}\right]^{1/2}_{}\\
\Lambda^{}_{+} & = &  \sqrt{\lambda(\lambda+\lambda^{}_{sd})/2}\; ,
\end{eqnarray}
\end{subequations}
\noindent will be used, over the whole range of $r$. 

The total mean free path of the electrons can be expressed in terms of the relative concentrations ($c$, and $c^{}_i$) of magnetic and non-magnetic impurities, and their respective scattering cross-sections ($\sigma^{}_{sd}$, and $\sigma^{}_{i}$) as
\begin{equation}
\lambda^{-1}_{}=n\sigma^{}_{sd}c+n\sigma^{}_i c^{}_i\; .
\end{equation}  
\noindent $n=4/a^{3}_{0}$ is the density of sites (in the fcc lattice, with lattice constant $a^{}_{0}$). It will be useful to write the $\sigma$s in terms of the resistivities ($\rho^{}_{sd}$, and $\rho^{}_{i}$). However, we must take into account that each of the latter  depends  on its corresponding transport cross-section\cite{Larsen}

\begin{equation}
\sigma^{\text{tr}}_{}=2\pi\int {\rm d}\theta(1-\cos\theta)\sin\theta\sigma(\theta)\; , 
\end{equation}
\noindent whereas the mean free path that takes place in the range of the interaction depends on the cross-section
\begin{equation}
\sigma=2\pi\int {\rm d}\theta\sin\theta\sigma(\theta)\; . 
\end{equation}
\noindent $\sigma(\theta)$ is the differential cross-section (one for each type of impurity). Employing the definition of electrical resistivity in metals, one can write $\sigma^{}_i=g^{}_i \rho^{}_i e^2_{}/ \hbar k^{}_{\rm F}$, and similarly for $\sigma^{}_{sd}$, where the ratios $g=\sigma/\sigma^{\text{tr}}_{}$ (one per impurity type) are left as adjustable parameters. A simple analysis, writing $\sigma(\theta)$ in terms of the partial waves phase-shifts (see \cite{Larsen}), shows that $1<g\sim 3$. 

A rather special type of effect has been observed in experiments discussed by Vier and Schultz (VS)\cite{VS}, related to changes in the concentration ($c^{}_{so}$) of  non-magnetic impurities with strong spin-orbit coupling. When $c^{}_{so}$ is increased (from zero), the interaction becomes more anisotropic; this  results in an initial increase for $T^{}_{\rm g}$, which then saturates because of the additional scattering caused by these impurities. The quantitative description of these effects is beyond  the scope of this work.
																									 
\subsection{The short distance cut-off $\boldsymbol{r^{}_0}$}
A simple choice would be to cut off the interaction at an average shortest distance. However, this arbitrary choice would exclude the possibility of spins occupying nearer neighbours sites; a situation that, although less probable, has a non-negligible contribution because of the power-law for $J(r)$. A more precise approach consists of introducing the probability distribution of nearest neighbour distance $\omega(r)$, so that
\begin{equation}
T^{}_{\rm g}=\int_{0}^{\infty} {\rm d}\xi \omega(\xi) T^{}_{\rm g}(\xi)\; , \label{Om1}
\end{equation}
\noindent where $ T^{}_{\rm g}(\xi)$ is given by the integral in  (\ref{CC0}) with $\xi$ as the lower integration limit.  $\omega(r)$ was derived by Chandrasekhar\cite{rmpstat}, for the 
continuum, and can be adapted to the disordered lattice problem as
\begin{equation}
\omega(r)=\left\{\begin{array}{ll}
n c' r^2_{}\exp[\frac{4\pi n c'}{3}(d^{3}_0-r^3_{})] & \text{for}\quad r\geq d^{}_0\\
       0 & \text{for}\quad r< d^{}_0\; ,
\end{array} \right.
\end{equation} 
\noindent where $d^{}_0=2^{-1/2}_{}a^{}_0$ is the nearest-neighbours distance in the fcc lattice, and $c'=c/(1-c)$  ensures the correct $c\rightarrow 1$ limit (in which $\omega(r)$ should become a delta function)\footnote{The correction $(1-c)$ is, in principle, unimportant at low concentrations.}.
 A similar approach can be found in \cite{Larsen,Larsen2}; where the authors took the  atomic radius $r^{}_{\text{at}}=a^{}_0 (3/16\pi)^{1/3}_{}$ as shortest possible distance. Here $d^{}_0$ is chosen instead,  because taking  $r^{}_0=r^{}_{\text{at}}\approx 0.55 d^{}_0$ would  count interactions between spins at un-physically close distances.
 
 One can switch the order of the integration in (\ref{Om1}), which gives
 
\begin{equation}
T^{}_{\rm g}=2\pi n c \int_{d^{}_0}^{\infty} r^2_{} {\rm d}r  \{1-\exp[4\pi n c'(d^{3}_0-r^3_{})/3]\} \overline{|G|}\; \overline{|J|}\; .\label{Pre}
\end{equation}  
\noindent The function $f(r)=n c \{1-\exp[4\pi n c'(d^{3}_0-r^3_{})/3]\}$  rapidly approaches a constant $f^{}_{\infty}=nc$ (the global density of spins) as $r$ grows. For  $r<d^{}_0$ it is identically zero. The volume integral of $[f^{}_{\infty}-f(r)]$ equals one; i.e., the excluded central spin.    The average volume for this  spin, to leading order in $c$, is $1/nc$, which provides the length scale inherent to the positionally disordered spin system $r^{}_c=(3/16\pi c)^{1/3}_{}a^{}_0$.
  
\subsection{Equation and fitting parameters} 
Having established the form of $J(r)$, and the relevant length scales, it is useful to define the dimensionless distance $x=r/d^{}_0$, which takes values in the range $[1;\infty)$. Doing this, the complete solution for $T^{}_{\rm g}$ is rewritten in as    

\begin{equation}
T^{}_{\rm g}=J^{}_0\int_1^{\infty}\frac{{\rm d}x}{x} [1-\text{e}^{b^{}_c(1-x^3_{})}_{}]h(x)\overline{|G|}(b^{1/3}_c x)\; , \label{final}
\end{equation}
\noindent where
\begin{eqnarray}
b^{}_c&=& b c \; ,\nonumber \\
b &=& \frac{4\pi\sqrt{2}}{3}\; , \nonumber \\
\overline{|G|}(t)&=& [\text{min}(1,1/t)]^{1+\eta'}_{}\; ,\nonumber  \\
h(t)&=&\left[\frac{13}{16}\text{e}^{-t z^{}_{sd}}_{}+\frac{3}{16}\text{e}^{-t z^{}_{T^{}_{\rm g}}}_{}\right]^{1/2}_{}\; , \label{aux}\\
\gamma &=&\frac{2 e^2_{}}{\pi\hbar a^{}_0}\; , \nonumber \\
z^{}_{sd} &=& \gamma b^{1/3}_{} (1+\rho^{}_{i}g^{}_{i}/2\rho^{}_{sd}g^{}_{sd})^{-1/2}_{} (\rho^{}_{sd}g^{}_{sd}+\rho^{}_{i}g^{}_{i}) \; ,\nonumber \\
z^{}_{T^{}_{\rm g}}&=& 2\pi(3 \gamma T^{}_{\rm g}/T^{}_{\rm F})^{1/2}_{}(\rho^{}_{sd}g^{}_{sd}+\rho^{}_{i}g^{}_{i})^{1/2}_{} \; , \nonumber \\
J^{}_0& = & \kappa 3 J^2_{sd} S^2_{}(2l+1)^2_{}/2\pi T^{}_{\rm F} \; .\nonumber
\end{eqnarray}

The function $\overline{|G|}(t)$ takes care of the fall-off of the correlations for $r>r^{}_{c}$ (note that $b^{-1/3}_c=r^{}_c/d^{}_0$). The variable $\kappa$, included in the multiplicative constant $J^{}_0$, is the numerical prefactor of the correlations, whose exact value is unknown. Thus, it is left as an adjustable parameter of the model. The degree of anisotropy in the interactions, which have not been considered so far, may also affect the value of $\kappa$\cite{BMY}. As  (\ref{aux}) contains interpolations between asymptotic expressions, one should be cautious while interpreting the best-fitting values for $\kappa$. It may be wise just to check whether it satisfies the physical condition $\kappa\lesssim 1$.  

The second  adjustable parameter of the model is $g^{}_{sd}$ (or $g^{}_{i}$), which has already been introduced in section \ref{jeff}. The experimental data will be presented in a way that the concentration of only one type of impurity is varied at a time. Varying $c$ first, with no other impurity present, will allows us to fix $g^{}_{sd}$. Once $g^{}_{sd}$ is fixed,  the behaviour of  $T^{}_{\rm g}$ as a function of the concentration of non-magnetic impurities (or the related resistivity) can be modelled, and the value of  $g^{}_i$ (corresponding to that type of impurity) can be found.  The critical exponent $\eta$ will also be optimized, in order to find the universality class (Ising-glass, Heisenberg-glass, or chiral-glass\cite{Kawamura}) that best describes the experiments.    
 
The integral (\ref{final}) is evaluated numerically using Boole's rule. The parameters $\kappa$, and $g$ are also found numerically,  employing a public routine LMDIF, which  minimizes the sum of the squares of non-linear functions in several variables, by a modification of the Levenberg-Marquardt algorithm\cite{minpack}.

\section{Experimental verification}\label{Exp}

Let us commence with the {\sl Au}Fe system. Larsen collected a considerable amount of data from the literature, which he tabulated in\cite{Larsen}. Only  17 samples, from which the electrical resistivity at the critical point  is also given, can be employed in the optimization of the parameters for our model. Perhaps, because of the differences in the preparation, these samples do not show a simple $\rho$ versus $c$ relation. Note that, as shown by  detailed studies\cite{Myd,resist} of the resistivity in these alloys (and also for {\sl Ag}Mn and {\sl Cu}Mn),   the temperature-dependent part of $\rho$  can only account for deviations of  about 5{\%} from the simple form $\rho(T^{}_{\rm g})\propto c$.    Without a  continuous form for the resistivity as a function of $c$, we can only calculate $T^{}_{\rm g}$ at the points where $\rho(T^{}_{\rm g})$ is given, and rather display the data in a scatter-plot.  The points with unknown $\rho(T^{}_{\rm g})$ cannot be included in the optimization. Assuming in our model that only iron is present, $z^{}_{sd}$ and $z^{}_{T^{}_{\rm g}}$ take the simple forms
\begin{subequations}
\begin{eqnarray}
z^{}_{sd} & = & \gamma b^{1/3}_{} \rho^{}_{sd}g^{}_{sd}\; , \label{mod1} \\
 z^{}_{T^{}_{\rm g}} & = & 2\pi(3 \gamma T^{}_{\rm g}/T^{}_{\rm F})^{1/2}_{}(\rho^{}_{sd}g^{}_{sd})^{1/2}_{} \; . \label{mod2}
\end{eqnarray}
\end{subequations}
\noindent Naturally, the value of $g^{}_{sd}$ will be influenced by all other unaccounted sources of scattering, which seem to be present in this case. 

\begin{figure}[htb]
\centerline{\includegraphics[width=0.55\linewidth]{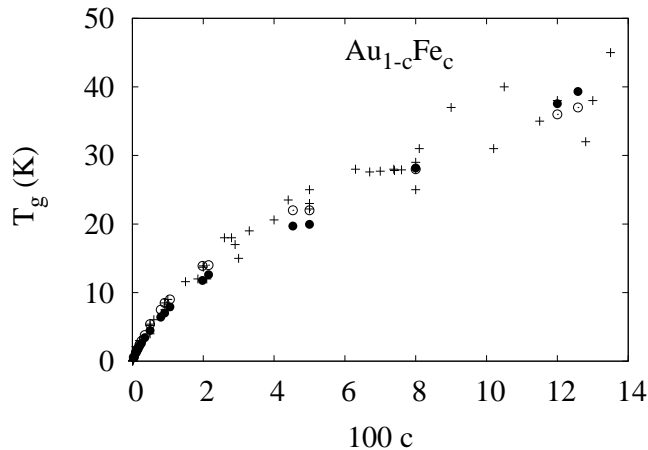}}
\caption{The freezing temperature ($T^{}_{\rm g}$) as a function of the concentration of magnetic (Fe) impurities in {\sl Au}Fe alloys. Data employed for the fit, from \cite{Larsen} (circles), other experimental points from \cite{Larsen} (crosses), and theoretical fit (black dots).}\label{Tfrho}
\end{figure}

In figure \ref{Tfrho}, $T^{}_{\rm g}$ is shown as a function of $c$. The circles represent the experimental data (of the 17 samples employed for the fit); the black dots, the  theoretical fit with $g^{}_{sd}=3.41$ and $\eta=-0.41$. The  points that were not included in the optimization are displayed as crosses.  Taking  $a^{}_0=4.08$ {\AA}, $T^{}_{\rm F}= 5.5$ eV, $k^{}_{\rm F}= 1.2$ {\AA}$^{-1}_{}$, $n=5.9$ $10^{28}_{}$ m${}^{-3}_{}$, $S=1.2$, and $J^{}_{sd}= 0.24$ eV \cite{Schilling}, we obtain $\kappa=0.32$. As such, all fitting parameters are in the appropriate range; the value for $g^{}_{sd}$ being similar to those discussed by Larsen$\cite{Larsen,Larsen2}$.  If  $\eta$ is fixed to a different value, the other fitting parameters, and the quality of the fit, differ little from the optimal values.  We shall come back to this point after analysing other materials.  

\begin{figure}[htb]
\centerline{\includegraphics[width=0.55\linewidth]{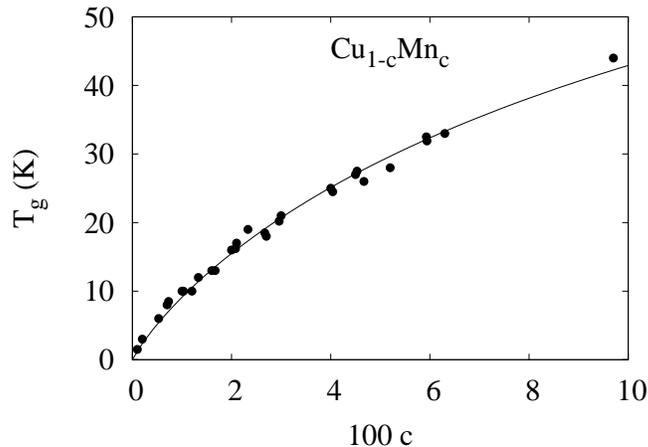}}
\caption{The freezing temperature $T^{}_{\rm g}$ in {\sl Cu}Mn alloys {\em vs} concentration of Mn. Experimental data from \cite{Myd,Myd2,Cowen,VS} (black dots), and theoretical fit (solid line).}\label{CuMnfit}
\end{figure}

Now we move to {\sl Cu}Mn alloys, employing the data from  Mydosh {\it et al} \cite{Myd,Myd2}, Cowen {\it et al} \cite{Cowen}, and VS \cite{VS}. The experimental points correspond to samples with  (nominally) only magnetic impurities (Mn).  For this system, one has $a^{}_0=3.61$ {\AA}, $n=8.5$ $10^{28}_{}$ m${}^{-3}_{}$,  $T^{}_{\rm F}= 7$ eV, $k^{}_{\rm F}= 1.36$ {\AA}$^{-1}_{}$, $S=1.95$, and $J^{}_{sd}= 0.21$ eV \cite{Schilling}. As reported elsewhere \cite{Cowen,Geldart2}, the resistivity in these alloys  can be roughly represented by the linear dependence  $\rho=400 c$ $\mu\Omega$ cm. 

The experimental data is shown as black dots in figure \ref{CuMnfit}; the solid line is the best fit for  (\ref{final}), with $g^{}_{sd}=1.79$, $\kappa=0.16$, and $\eta=-0.41$. Here we draw a line for the calculated values, instead of isolated points, in accord with the information available for $\rho(T^{}_{\rm g})$. The theory reproduces the experiments over a wide range of concentrations (from $c=0.0001$ to $c=0.1$). While the fall-off of the correlations is important in the description of the low concentrations regime (interaction with very long range), the success of the theory in the higher-$c$ region depends on the treatment of the exponential tail of the interaction. Note the weaker decrease of $r^{}_c\sim c^{-1/3}_{}$, in comparison with $\lambda\sim c^{-1}_{}$, and $\Lambda^{}_T\sim (cT)^{-1/2}_{}$. An alternative approach, where the interaction is sharply cut in its range\cite{Geldart2}, could not reproduce the experiments  as well as it is shown in figure \ref{CuMnfit}, over the whole $c$ domain.  When the range of the interaction becomes similar to the typical distance between the spins, the major contribution to $T^{}_{\rm g}$ comes from the exponential tail, which is lost if one takes the sharp-cut-off approach.

As last example, we examine the data from {\sl Ag}Mn alloys reported by Vier and Schultz (VS) \cite{VS}. VS studied both cases: case (a) where manganese is the only impurity and its concentration is varied; case (b) where the number of magnetic (Mn) impurities is fixed (at $c=0.026$),  and non-magnetic Sb-impurities are added. For these alloys, one finds in the literature $a^{}_0=4.09$ {\AA}, $n=5.86$ $10^{28}_{}$ m${}^{-3}_{}$, $T^{}_{\rm F}= 5.5$ eV, $k^{}_{\rm F}= 1.2$ {\AA}$^{-1}_{}$, $S=1.9$,  and $J^{}_{sd}= 0.18$ eV \cite{Schilling}.

\begin{figure}[htb]
\centerline{\includegraphics[width=0.55\linewidth]{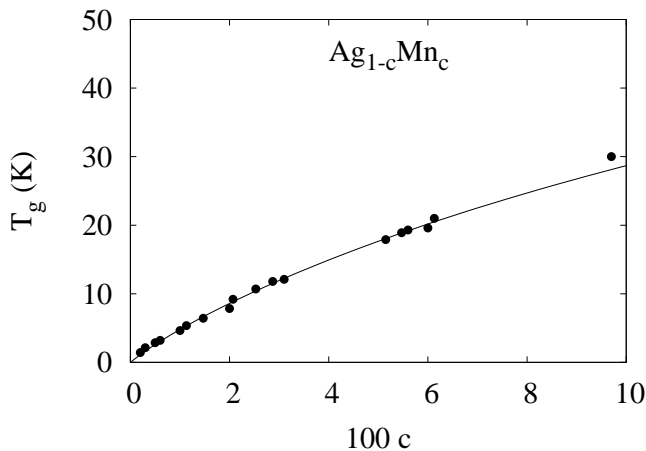}}
\caption{The freezing temperature ($T^{}_{\rm g}$) in {\sl Ag}Mn alloys {\em vs}  concentration of Mn. Experimental data from \cite{VS} (black dots), and theoretical fit (solid line).}\label{VSrep}
\end{figure}

For case (a),  $z^{}_{sd}$   and $z^{}_{T}$ take the forms (\ref{mod1}) and (\ref{mod2}), respectively. The resistivity is modelled as $\rho=153.8 c$ $\mu\Omega$ cm \cite{Geldart2}. The experimental points collected from the literature are shown in   figure \ref{VSrep}, as black dots. The best-fitting curve, in the range $0<c<0.1$, is shown as a solid line.  The fitting  parameters are $g^{}_{sd}=1.73$, and $\kappa=0.08$\footnote{This value for $g^{}_{sd}$ is slightly different to the one obtained for {\sl Cu}Mn ($g^{}_{sd}=1.79$).  Although {\sl Ag}Mn and {\sl Cu}Mn share the same impurity type, the $g^{}_{sd}$s may differ on account of a different environment (hybridization, Fermi energy, etc). See, for example, similar effects on the $s$-$d$ couplings ($J^{}_{sd}$) \cite{Schilling}. The differences in $g^{}_{sd}$ may also be due to the approximated character of the theory.}.  

Here, as well as with {\sl Cu}Mn, and {\sl Au}Fe, the quality of the fit worsens slightly as one forces $\eta$ away from the Ising exponent.   It should be noted that if, instead of using (\ref{aux}), one models the interaction cutting it sharply at  $\Lambda^{}_{T^{}_{\rm g}}$, the dependence of on $\eta$ is much stronger. In that case, only the Ising exponent permits  acceptable fits. The interpolation (\ref{aux}), and the sharp cut-off at $\Lambda^{}_{T^{}_{\rm g}}$ are two extreme approaches to an intermediate,  unknown, functional dependence. However,  both choices lead to the same conclusion. These results are in agreement with the work of Bray {\it et al} \cite{BMY}, where they concluded that ``all three-dimensional experimental spin-glass systems should have a transition in the universality class of the short-range Ising spin-glass model''.

\begin{figure}[htb]
\centerline{\includegraphics[width=0.55\linewidth]{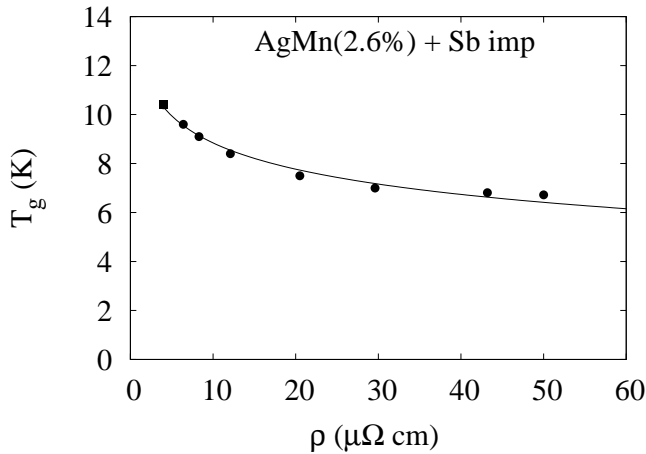}}
\caption{The freezing temperature ($T^{}_{\rm g}$) in {\sl Ag}Mn alloys {\em vs} total resistivity, with Sb impurities and 2.6 at.{\%}Mn. Experimental data from \cite{VS} (black dots), and theoretical fit (solid line).}\label{VSrep2}
\end{figure}

Having fixed $g^{}_{sd}$ for Mn, we can analyse situation (b).  $z^{}_{sd}$   and $z^{}_{T}$ take the more general forms given in  (\ref{aux}), with $g^{}_{sd}=1.73$. Assuming that changes in spin-orbit effects from Sb impurities are not strong, one can use the  value for  $\kappa$ previously obtained. Then, there is  a single free parameter, $g^{}_i$(Sb), to adjust the model to the new data. The freezing temperature, as a function of the total resistivity\footnote{According to VS, the resistivity is well described by the linear dependence $\rho=4+740c^{}_i$ $\mu\Omega$ cm, where $c^{}_i$ is the concentration of Sb-impurities \cite{VS}.} is shown in figure \ref{VSrep2}: experimental points, as dots; theory with $g^{}_i=1.68$, as a solid line. The black square corresponds to the point with only Mn-impurities.  If one also takes  $\kappa$ as a  free parameter,  one finds no significant changes; neither in $\kappa$, nor in $g^{}_i$.

Equation (\ref{final}) describes acceptably the decrease of   $T^{}_{\rm g}$ with the addition of non-magnetic impurities. As indicated by SG \cite{Geldart2}, the flattening tendency is a consequence of a slower reduction of the effective range in comparison with the total mean free path. Nevertheless, the experimental points seem to have a tendency toward  a total saturation, which is not indicated by the theory. VS found  that their data are well described by the experimental formula
\begin{equation}
T^{}_{\rm g}(\rho)=T^{}_{\rm g}(\infty) +(T^{}_{\rm g}(0)-T^{}_{\rm g}(\infty))\text{e}^{-\rho/\rho^{}_0}_{}\; , \label{experVS}
\end{equation}
\noindent with $\rho^{}_0=10$ $\mu\Omega$ cm, $T^{}_{\rm g}(\infty)=6.7$ K, and $T^{}_{\rm g}(0)=12.4$ K. One can show that, because of the strong behaviour of the exponential function, the data could also be described by more complicate functions of $\rho$ in its argument. From the perspective of the present theory, using the mean-value theorem, the leading order of   (\ref{final}) in $\rho$ contains exponentials with sub-linear dependences on $\rho$. However the non-zero asymptotic value   $T^{}_{\rm g}(\infty)$ remains unexplained. 

To understand this, we should go back to a very fundamental problem, the  mean free path of the electrons. Most intentions of describing the RKKY interaction, in the presence of disorder, have been carried out under the assumption that $\lambda k^{}_{\rm F}\gg 1$. Also, the standard definition of mean free path is meaningful as long as its calculated value is reasonably larger than the average  distance ($\overline{d}$) between the scatterers.  In the high-$\rho$ regime presented in figure (\ref{VSrep2}),  values of $\lambda k^{}_{\rm F}\sim 10$, and $\lambda/\overline{d}\sim 3$ are reached.  When $\lambda$ is reduced to values of the order of the distance between scatterers, deviations from the $\lambda\propto \rho^{-1}_{}$ behaviour should be expected. Thus, in order to account for the physics at very high-$\rho$, we require a (new) theory of the RKKY interaction beyond the weak scattering regime. Larsen showed \cite{Larsensat} that the saturation of $T^{}_{\rm g}$ can be reproduced within the MFF employing  
\begin{equation}
\lambda^{-1}_{}=z\rho [1+(\rho/\rho^{}_0)^2_{}]^{-1/2}_{}\label{Kane}
\end{equation}
\noindent (where $z$, and $\rho^{}_0$ are material dependent phenomenological constants). Certainly, such a functional form would also allow us a perfect fit with  (\ref{final}), giving $T^{}_{\rm g}\propto \text{e}^{-\gamma \rho [1+(\rho/\rho^{}_0)^2_{}]^{-1/2}_{} }_{}$, to leading order in $\rho$ ($\gamma$ is a mean-value constant). However, although (\ref{Kane}) is qualitatively reasonable, it has only been derived \cite{LarsenKane} from the theory  of Kaneyoshi \cite{KaneTheo1,KaneTheo2}, which relies on special assumptions about the energy dependence of the one-electron Green function. Up to date, it is not clear whether such a dependence is completely physical or not \cite{D2}.

\section{Summary and outlook}\label{End}
To conclude, we have derived an equation for the determination of  the freezing temperature ($T^{}_{\rm g}$) in canonical spin-glasses, which  gives a coherent description of its concentration dependence, and reproduces multiple experiments well. The analytical derivation was based on the statistical properties of the couplings  and  correlations, and did not require an average over disorder.   The  price to pay for including the correlations is having an unknown numerical prefactor, related to the specific shape of their distribution; i.e., the asymptotic form of the correlations is know analytically, $G(r)\propto r^{-1-\eta'}_{}$, to within a proportionality constant.   It would be interesting to test the present results with numerical simulations, where correlations can be computed exactly.  Since this power-law scaling  takes part in the solution for $T^{}_{\rm g}$, information about the critical exponent can be obtained from the comparison with experiments. The Ising universality class is found to have the best fit to the  data. This is plausible since the fixed point of the isotropic Heisenberg model  is unstable with respect to anisotropies in the coupling constants, and the long-ranged interaction in metals is known to have a certain degree of anisotropy.   Nevertheless, the sensitivity of the method, to changes in the critical exponent, depends on how one approximates the exponential tail of the interaction.   Performing more measurements of $T^{}_{\rm g}$, and the electrical resistivity $\rho(T^{}_{\rm g})$ at low concentration of impurities  could provide us with more conclusive evidence. Finally,  effects beyond the weak scattering regime (not yet considered) seem to be important in the quantitative description of the $T^{}_{\rm g}$  versus $\rho$ dependence, observed in experiments with very dirty samples.

The author thanks professor Mark R. A. Shegelski for the stimulating correspondence. This work has been carried out in the group of Prof. R. Nesper, supported by the Swiss National Science Foundation through grant number 2-77937-10.

\end{document}